# LETTER

# A 17-billion-solar-mass black hole in a group galaxy with a diffuse core

Jens Thomas[1,2], Chung-Pei Ma[3], Nicholas J. McConnell[4], Jenny E. Greene[5], John P. Blakeslee[4] & Ryan Janish[6]

Quasars are associated with and powered by the accretion of material onto massive black holes; the detection of highly luminous quasars with redshifts greater than $z = 6$ suggests that black holes of up to ten billion solar masses already existed 13 billion years ago[1]. Two possible present-day 'dormant' descendants of this population of 'active' black holes have been found[2] in the galaxies NGC 3842 and NGC 4889 at the centres of the Leo and Coma galaxy clusters, which together form the central region of the Great Wall[3]—the largest local structure of galaxies. The most luminous quasars, however, are not confined to such high-density regions of the early Universe[4,5]; yet dormant black holes of this high mass have not yet been found outside of modern-day rich clusters. Here we report observations of the stellar velocity distribution in the galaxy NGC 1600—a relatively isolated elliptical galaxy near the centre of a galaxy group at a distance of 64 megaparsecs from Earth. We use orbit superposition models to determine that the black hole at the centre of NGC 1600 has a mass of 17 billion solar masses. The spatial distribution of stars near the centre of NGC 1600 is rather diffuse. We find that the region of depleted stellar density in the cores of massive elliptical galaxies extends over the same radius as the gravitational sphere of influence of the central black holes, and interpret this as the dynamical imprint of the black holes.

We observed NGC 1600 (Fig. 1) as part of the MASSIVE Survey[6], the aim of which is to study the structure, dynamics, and formation history of the 100 most massive early-type galaxies within 108 megaparsecs (Mpc) of Earth. This volume-limited survey probes galaxies with stellar masses above $5 \times 10^{11} M_\odot$ (where $M_\odot$ is the mass of the Sun) in diverse, large-scale environments that have not been systematically studied before. The stellar mass ($8.3 \times 10^{11} M_\odot$), halo mass ($\sim 1.5 \times 10^{14} M_\odot$), and distance (64 Mpc) of NGC 1600 are fairly typical of the galaxies in the survey. We obtained stellar spectra covering the central 5-arcsec-by-7-arcsec region of NGC 1600 with roughly 0.6-arcsec spatial resolution, using the Gemini multi-object spectrograph (GMOS) at the Gemini North Telescope. We further obtained large-area (107-arcsec-by-107-arcsec) stellar spectra of NGC 1600 using the Mitchell integral field spectrograph (IFS) at the McDonald Observatory. The stellar luminosity distribution of the galaxy is provided by surface photometry from the Hubble Space Telescope (HST) and the Kitt Peak National Observatory[7].

We measured the distribution of the line-of-sight stellar velocities at 86 locations in NGC 1600 by modelling the deep calcium triplet absorption lines in our GMOS IFS spectra and several optical absorption features in our Mitchell IFS spectra. The galaxy shows little rotation (less than $30 \text{ km s}^{-1}$), and the line-of-sight velocity dispersion rises from 235–275 $\text{km s}^{-1}$ at large radii to a maximum value of 359 $\text{km s}^{-1}$ near the centre, consistent with previous long-slit measurements[8].

We used orbit superposition models[9] to determine the mass of the central black hole, $M_{BH}$, of NGC 1600: we find a value for $M_{BH}$ of $(1.7 \pm 0.15) \times 10^{10} M_\odot$ (68% confidence interval). We rule out statistically the possibility that the $M_{BH}$ is less than $10^{10} M_\odot$ with more than 99.9% confidence (see Methods). Defining the sphere of influence of the black hole, $r_{SOI}$, as the radius at which the enclosed stellar mass equals $M_{BH}$, we find $r_{SOI} = 3.8$ arcsec (or 1.2 kiloparsec, kpc) for NGC 1600, using our measured $M_{BH}$ and our calculated value for $M_\star/L$ of $(4.0 \pm 0.15) M_\odot/L_\odot$ (in the R-band; $L_\odot$ is the solar luminosity). Our velocity data resolve the central spatial region of NGC 1600 down to about 200 pc, where $M_{BH}$ exceeds the enclosed stellar mass by a factor of 100. Even if the unresolved stellar mass near the centre were ten times larger because of, for example, an extreme population of undetected dwarf stars, this would not have a measurable effect on $M_{BH}$ (see Methods).

The stellar bulge mass of NGC 1600 is $8.3 \times 10^{11} M_\odot$, according to our dynamical modelling, and is consistent with the value inferred from the absolute K-band magnitude of $-25.99$ for NGC 1600 (ref. 6). The $M_{BH}$ of NGC 1600 is 2.1 per cent of its bulge mass ($M_{bulge}$)—three to nine times more than the percentage predicted from the known scaling relations of black-hole and galaxy bulge mass[10–12]. Other galaxies with high $M_{BH}$-to-$M_{bulge}$ ratios are all notable for their compact size, suggesting that they are tidally stripped objects or relics from the young Universe with stunted late-time growth. In contrast, NGC 1600 is much less compact, as judged by its half-light radius, which is comparable to that of other giant elliptical galaxies of similar luminosity. The $M_{BH}$ of NGC 1600 is also about ten times larger than the mass expected from its average stellar velocity dispersion (Extended Data Fig. 1).

NGC 1600 has a remarkably faint and flat core: it has the largest core radius detected in a study of 219 early-type galaxies using HST

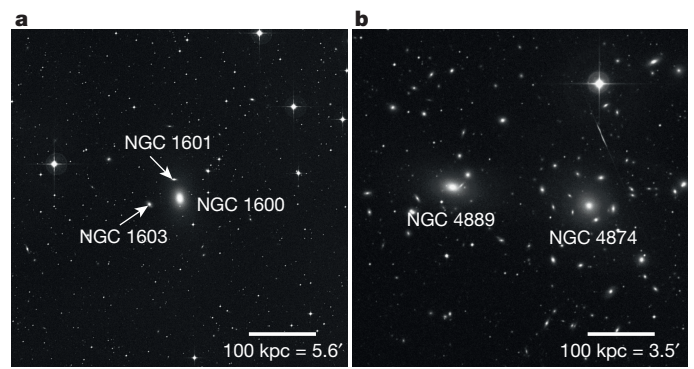

**Figure 1 | Environment of NGC 1600 versus that of NGC 4889. a**, The central 500 kpc × 500 kpc of the NGC 1600 group of galaxies; this group has a total mass[24,25] of about $1.5 \times 10^{14} M_\odot$ and an X-ray luminosity[26] of $3 \times 10^{41} \text{ erg s}^{-1}$. The two closest companion galaxies of NGC 1600 (NGC 1601 and NGC 1603) are nearly eight times fainter than NGC 1600. **b**, The innermost 500 kpc × 500 kpc of the Coma Cluster, which contains more than 1,000 known galaxies, and is at least ten times more massive[27,28] than the NGC 1600 group, and 1,000 times more X-ray-luminous[29]. NGC 4889 has twice the stellar mass of NGC 1600 and a nearly equally luminous neighbour galaxy (NGC 4874). Both images are from the Second Palomar Observatory Sky Survey (R-band; north is at the top, and east to the left).

[1]Max Planck-Institute for Extraterrestrial Physics, Giessenbachstraße 1, D-85741 Garching, Germany. [2]Universitätssternwarte München, Scheinerstraße 1, D-81679 München, Germany. [3]Department of Astronomy, University of California, Berkeley, California 94720, USA. [4]Dominion Astrophysical Observatory, NRC Herzberg Institute of Astrophysics, Victoria, British Columbia V9E 2E7, Canada. [5]Department of Astrophysical Sciences, Princeton University, Princeton, New Jersey 08544, USA. [6]Department of Physics, University of California, Berkeley, California 94720, USA.

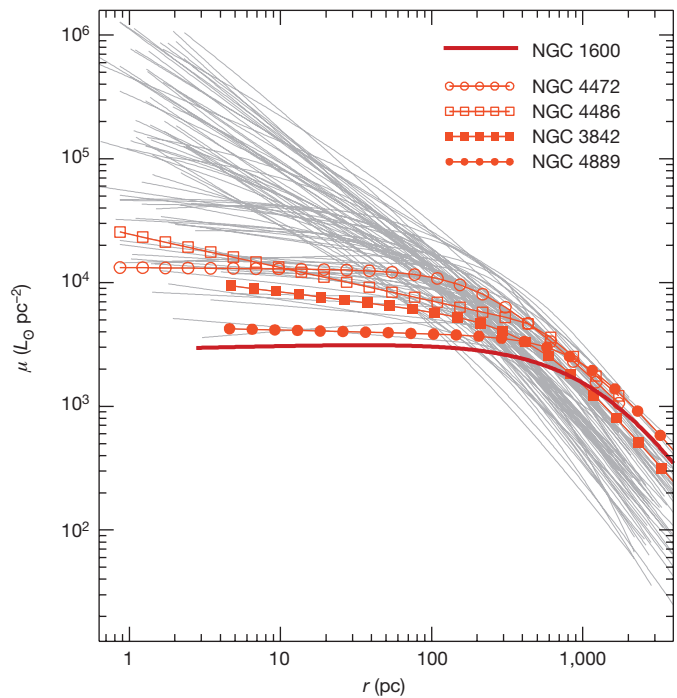

**Figure 2 | Central stellar light profiles for NGC 1600 and for a sample of other core and coreless elliptical galaxies.** Surface brightness profiles (in the V-band) are shown for a sample of galaxies, on the basis of HST observations[13] up to a distance of 100 Mpc from Earth. NGC 4889, at 102 Mpc, is included because its black-hole mass has been measured[2]. $\mu$ is the surface brightness and $r$ is the galactic radius at which the brightness was measured. Lower-luminosity elliptical galaxies typically have rising light profiles towards the galactic centres (steep grey curves), whereas NGC 1600 and other very massive elliptical galaxies often exhibit a marked deficit of stars in the central region (red curves). Highlighted are the brightest galaxies in the Leo Cluster (NGC 3842) and the Coma Cluster (NGC 4889), and the brightest (NGC 4472) and central (NGC 4486 or M87) galaxies of the Virgo Cluster. The stellar core in NGC 1600 (dark red curve) is the faintest known among all galaxies for which dynamical $M_{BH}$ measurements are available.

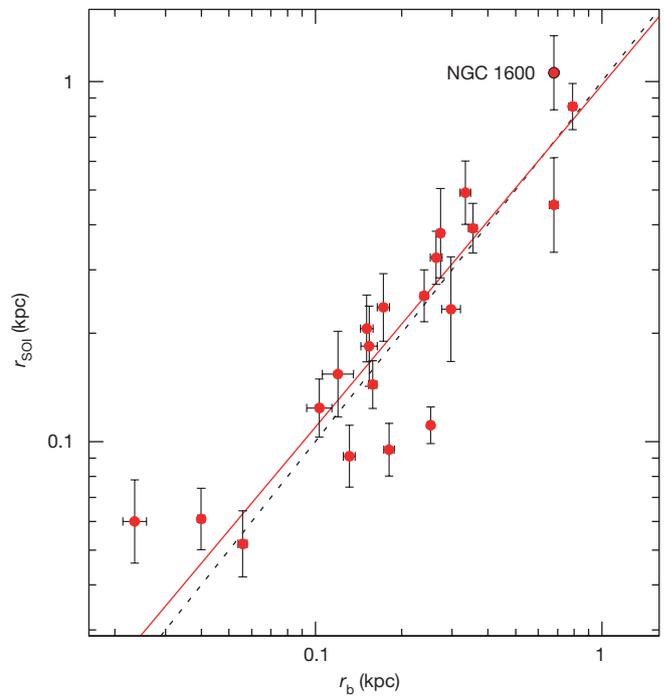

**Figure 3 | Black-hole sphere-of-influence radii and galaxy core radii.** The radius of the black-hole sphere-of-influence, $r_{SOI}$, of NGC 1600 and 20 core galaxies with dynamical $M_{BH}$ measurements[15] is plotted against the core radius, $r_b$, of each host galaxy. We calculate $r_{SOI}$ directly from the measured $M_{BH}$ and the de-projected stellar density profile of each galaxy. We calculate the core radius from fits to a core-Sérsic function[30]. Our best-fit linear correlation is $\log_{10}(r_{SOI}/\text{kpc}) = (-0.01 \pm 0.29) + (0.95 \pm 0.08)\log_{10}(r_b/\text{kpc})$ (straight red line), which is statistically consistent with $r_b = r_{SOI}$ (dotted black line). The intrinsic scatter of this relation is $\epsilon = 0.17 \pm 0.04$.

observations[13], and a lower central surface brightness than any other galaxy within 100 Mpc of Earth in this sample (3,100$L_\odot$ pc$^{-2}$; Fig. 2). Such a diffuse light distribution indicates a substantial deficit of stars in the central region of NGC 1600 in comparison with lower-luminosity elliptical galaxies, which typically show rising light profiles towards the galactic centres down to the smallest observable radius. Like NGC 1600, other very massive elliptical galaxies often exhibit a cored light profile, where the steep light distribution characteristic of the outer part of the galaxy flattens to a nearly constant surface brightness at small radius (Fig. 2). A plausible mechanism for creating depleted stellar cores is via three-body gravitational slingshots that scatter stars passing close to a supermassive black-hole binary to larger radii. While the stars are being scattered, the orbit of the black-hole binary shrinks and the binary will emit gravitational waves if it coalesces[14].

In Fig. 3 we present a new correlation between the radius, $r_b$, of the galaxy core in the observed light profile and the radius of the black hole's sphere of influence, for NGC 1600 and for a sample of 20 other core galaxies with reliable $M_{BH}$ measurements[15]. We find a tight connection between the two radii, and the best-fit relation is consistent with $r_b = r_{SOI}$. The intrinsic scatter of 0.17 dex in the $r_b$–$r_{SOI}$ relation is a factor of two smaller than that in the known scaling relations between black-hole mass and galaxy properties. This small scatter holds over a wide range of galaxy environments sampled in the 21 core galaxies. Our finding that the two radii, $r_b$ and $r_{SOI}$, are statistically indistinguishable with a small scatter strongly indicates that the core-formation mechanism is homogeneous and is closely connected to the central black hole. Gravitational core scouring by black-hole binaries can consistently produce the observed homology in the light profiles of galaxy cores[16,17] and the velocity anisotropy in stellar orbits[18] (see Methods).

Figure 4 shows the correlation between the core radius, $r_b$, and the measured $M_{BH}$ for NGC 1600 and the same sample of galaxies as is shown in Fig. 3. We find an intrinsic scatter of 0.3 dex in the $M_{BH}$–$r_b$ relation for these 21 core galaxies, in comparison to a root-mean-squared scatter of 0.41–0.44 dex in the $M_{BH}$–$\sigma$ relation (where $\sigma$ is the stellar velocity dispersion) for the same sample (Extended Data Fig. 1). The $M_{BH}$ values in core galaxies generally do not correlate well with the stellar velocity dispersion[10,11,19]. Simulated mergers of elliptical galaxies also suggest that the $M_{BH}$–$\sigma$ correlation may steepen or disappear altogether at the high-mass end[20]. The most massive black holes discovered thus far predominantly reside at the centres of massive galaxies containing stellar cores. For this high-mass regime, our findings indicate that the core radius of the host galaxy is more robust than its velocity dispersion as a proxy for $M_{BH}$.

The Schwarzschild radius (or 'event horizon') of the black hole in NGC 1600 is $5 \times 10^{10}$ km, or 335 AU, subtending an angle of 5.3 µas on the sky (at a distance of 64 Mpc). Recent mass measurements[21,22] of the black hole in the galaxy M87 range from $3.3 \times 10^9 M_\odot$ to $6.2 \times 10^9 M_\odot$, corresponding to a Schwarzschild radius of 3.8–7.3 µas (at 16.7 Mpc). For comparison, the Schwarzschild radius of Sagittarius A* is 10 µas. Thus, after the Milky Way and possibly M87, NGC 1600 contains the next most easily resolvable black hole, and is a candidate for observations with the Event Horizon Telescope[23].

Black holes with masses of about $10^{10} M_\odot$ are observed as quasars in the young Universe[1]. Finding the dormant-black-hole descendants of these luminous quasars and understanding their ancestral lineages have been strong motivations for the search for very massive black holes nearby. At redshifts of 2 to 3 (about 10 billion years ago), when the quasar activity peaks, the luminosities of the brightest quasars (more than $10^{48}$ erg s$^{-1}$) suggest the existence of very massive black holes (with masses of more than $5 \times 10^9 M_\odot$). Yet, on average, these quasars do not appear to inhabit environments that are notably different from those of their less luminous counterparts[4,5]. The 17-billion-solar-mass black hole in NGC 1600 is a possible descendant of a luminous quasar in an environment outside the richest structures of galaxies. Very massive dark-matter haloes are therefore not necessary for growing very massive black holes and their luminous quasar progenitors. The number density of dark-matter haloes of comparable mass to that of NGC 1600 is about $2 \times 10^{-5}$ Mpc$^{-3}$, making such haloes about 50 times more abundant than rich structures like the Coma Cluster. At present, we do not know whether dormant black holes with masses greater than $10^{10} M_\odot$ are common in other nearby $\sim 10^{14} M_\odot$ haloes with massive galaxies, or whether other properties of the NGC 1600 group—such as the large magnitude gap, fossil-group-like evolved dynamical state[24], and isolation—are necessary ingredients for cultivating such massive black holes. Our ongoing observations of massive galaxies[6] will soon reveal whether the extreme black hole in NGC 1600 is a rare find in an unusual environment, or the tip of an iceberg.

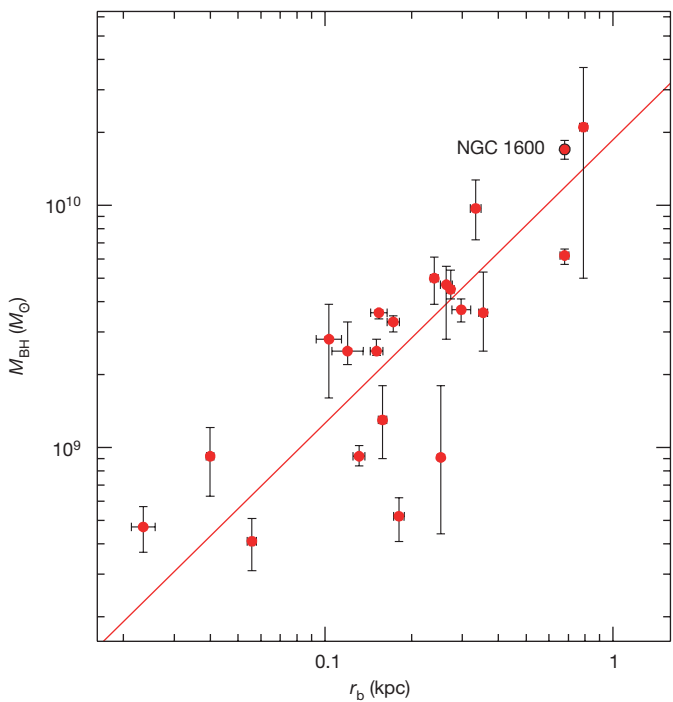

**Figure 4 | Black-hole mass and galaxy core radius.** The black-hole mass, $M_{BH}$, is plotted against the core radius, $r_b$, for the same sample of galaxies as in Fig. 3. The straight line shows our best fit for a constant-exponent power-law relation between $M_{BH}$ and $r_b$: $\log_{10}(M_{BH}/M_\odot) = (10.27 \pm 0.51) + (1.17 \pm 0.14)\log_{10}(r_b/\text{kpc})$. The intrinsic scatter of this relation is $\epsilon = 0.29 \pm 0.07$.

**Online Content** Methods, along with any additional Extended Data display items and Source Data, are available in the online version of the paper; references unique to these sections appear only in the online paper.

Received 28 October 2015; accepted 29 January 2016.
Published online 6 April 2016.

**Acknowledgements** C.-P.M., J.E.G. and R.J. are supported by the National Science Foundation (NSF). J.E.G. is supported by the Miller Institute for Basic Research in Science, University of California, Berkeley. N.J.M. is supported by the Beatrice Watson Parrent Fellowship and Plaskett Fellowship. The spectroscopic data presented here were obtained from the Gemini Observatory and the McDonald Observatory. Gemini is operated by the Association of Universities for Research in Astronomy, Inc., under a cooperative agreement with the NSF on behalf of the Gemini partnership. The McDonald Observatory is operated by the University of Texas at Austin. The photometric data presented here are based partly on observations made with the NASA/ESA Hubble Space Telescope, and obtained from the Hubble Legacy Archive, which is a collaboration between the Space Telescope Science Institute (STScI/NASA), the Space Telescope European Coordinating Facility (ST-ECF/ESA) and the Canadian Astronomy Data Centre (CADC/NRC/CSA).

**Author Contributions** J.T. developed and carried out the stellar orbit modelling. J.T. and C.-P.M. wrote the manuscript. C.-P.M. led the Gemini observation proposal. J.E.G. performed the stellar population analysis. N.J.M. and R.J. reduced the spectroscopic data. J.P.B. provided photometric analysis. All authors contributed to the MASSIVE Survey, the kinematic extractions, the interpretive analysis of the observations and the writing of the paper.

**Author Information** Reprints and permissions information is available at www.nature.com/reprints. The authors declare no competing financial interests. Readers are welcome to comment on the online version of the paper. Correspondence and requests for materials should be addressed to J.T. (jthomas@mpe.mpg.de) or C.-P.M. (cpma@berkeley.edu).

## METHODS

In the first section below, we describe spectroscopic data for NGC 1600 and our procedures for measuring stellar kinematics. In the second section, we describe photometric data and the surface brightness profile of NGC 1600. In the third section, we describe our stellar orbit modelling procedures. In the fourth section we compare the stellar mass-to-light ratio obtained from the dynamical models with independent constraints from a stellar population analysis. The implied distribution of stellar orbits in NGC 1600 is described in the fifth section. Finally, in the last section we present the black-hole scaling relations for core radii, determined from an alternative fit to the light profiles.

**Spectroscopic data and stellar kinematics.** We obtained high-spatial-resolution spectroscopic data from GMOS-N, an IFS on the 8-metre Gemini North Telescope. We observed the central region of NGC 1600 with GMOS-N, which provides continuous two-dimensional coverage of a 5 arcsec × 7 arcsec science field and simultaneously covers a sky field offset by 60 arcsec. Our spectra were centred on the triplet of calcium absorption lines from 8,480 Å to 8,680 Å, a well studied region frequently used for stellar kinematic measurements[31,32]. We obtained nine 1,230-second exposures of NGC 1600 over three nights of queue-mode observations in November 2014.

GMOS-N is seeing-limited, with spatial sampling of 0.2 arcsec on the IFS. We estimated the point-spread function (PSF) on each observing night by measuring the width of foreground stars in the acquisition images of NGC 1600. Our average PSF for GMOS-N has a full width at half-maximum (FWHM) of 0.6 arcsec. A Gaussian model of the PSF is included in our orbit superposition models.

We used the image reduction and analysis facility (IRAF) software package supplied by the Gemini Observatory to flat-field and wavelength-calibrate the GMOS-N data, and to extract a one-dimensional spectrum for each IFS lenslet. We developed custom routines to construct collapsed images of the galaxy and record the position of each one-dimensional spectrum with respect to the galaxy centre. The spectra were then resampled on a two-dimensional grid and binned to a consistent signal-to-noise ratio (of about 100 per pixel) using Voronoi tessellation[33]. Our binning implementation imposed symmetry over four projected quadrants of the galaxy, so that the kinematic measurements could be folded into a single quadrant before orbit modelling.

We obtained wide-field spectroscopic data from the Mitchell Spectrograph[34] on the 2.7-metre Harlan J. Smith Telescope at the McDonald Observatory. The Mitchell Spectrograph is an optical IFS with a 107 arcsec × 107 arcsec field of view and 246 fibres, each of 4.1-arcsec diameter. The low-resolution blue setting ($R \approx 850$) was used, providing wavelength coverage from 3,650 Å to 5,850 Å, including the Ca H+K region, the G-band region, H$\beta$, the Mg$b$ region, and several Fe absorption features. The spectral resolution varied spatially and with wavelength, with an average of 5 Å FWHM, corresponding to a dispersion of $\sim$1.1 Å pixel$^{-1}$ and $\sigma_{inst} \approx 100$ km s$^{-1}$ in the red and $\sigma_{inst} \approx 150$ km s$^{-1}$ in the blue part of the spectrum (where $\sigma_{inst}$ is the instrumental resolution). Data reduction was performed using the Vaccine package[35]. We fit Mitchell spectra with the MILES library of 985 stellar spectra[36] and determined the best-fit line-of-sight velocity distribution (LOSVD) for each of the 58 spatial bins[6,37]. Our GMOS and Mitchell data are the first IFS observations of NGC 1600 and reveal a stellar velocity distribution that is well aligned with the galaxy's light distribution, indicating that NGC 1600 is axisymmetric.

**Photometric data and core-Sérsic fit.** We used a high-resolution image taken with the near-infrared camera and multi-object spectrometer (NICMOS) instrument on the HST to measure the central light distribution of NGC 1600. The observation (from General Observer Program number 7886) consisted of four dithered exposures of NGC 1600 taken with NICMOS camera 2 in the F160W bandpass. We downloaded the calibrated, combined image from the Hubble Legacy Archive. The image had a pixel scale of 0.05 arcsec per pixel and total exposure time of 460 seconds.

We combined the HST observations with ground-based photometric data at large radii taken from the literature[7]. The NICMOS data were calibrated to the R-band of the ground-based data by minimizing the squared magnitude differences between the two surface brightness profiles in the radial region where both data sets overlap and PSF effects are negligible ($r = 2$–10 arcsec). Single one-dimensional profiles of the surface brightness, the ellipticity and the isophotal shape parameters[38] $a_4$ and $a_6$ were then constructed by using the NICMOS data at $r < 8$ arcsec and the ground-based data at $r \geq 8$ arcsec. The position angle (PA) of the isophotes is constant with radius[7] ($\Delta$PA < 2°), consistent with an axisymmetric stellar distribution.

The resulting circularized surface brightness distribution of NGC 1600 is well described by a core-Sérsic function with a core radius of $r_b = 2.15$ arcsec (Extended Data Fig. 2). Over several orders of magnitude in radius, the light profiles of lower-luminosity elliptical galaxies follow a single Sérsic function characterized by the Sérsic index $n$, the half-light radius $r_e$ and a surface brightness scale $\mu_e = \mu(r_e)$. The core-Sérsic function combines a Sérsic profile at $r > r_b$ and a power-law distribution with slope $\gamma$ at $r < r_b$. The transition is controlled by a smoothness parameter $\alpha$ and the surface brightness scale is $\mu_b = \mu(r_b)$. Inside $r < 5$ arcsec, the inward extrapolation of the outer Sérsic component overpredicts the central surface brightness of NGC 1600 by about three magnitudes. From the difference between the integrated light of the Sérsic component and the actual core-Sérsic fit, we derive a 'light deficit' of $\Delta L_{def} = 9.47 \times 10^9 L_\odot$ in the centre of NGC 1600.

We use the isophotal model of NGC 1600 to compute the galaxy's intrinsic luminosity density distribution. The deprojection is nonparametric[39] and accounts for the observed ellipticity profile and boxy shape of NGC 1600's isophotes[7]. The same technique has been used for the dynamical modelling of other galaxies[12,40].

**Stellar orbit models.** We generate dynamical models of NGC 1600 using Schwarzschild's orbit superposition method[41]. Because the two-body relaxation time of stars in massive elliptical galaxies exceeds the age of the Universe, their dynamics is governed by the collisionless Boltzmann equation. Any steady-state equilibrium solution of the Boltzmann equation can be written as a sum over single-orbit distribution functions, where the phase-space density of stars along each trajectory is constant. The total number of stars on each orbit—that is, the orbital luminosity or orbital occupation number—can take arbitrary (positive) values.

For the modelling, we assume that NGC 1600 is axisymmetric and that the stellar mass profile follows the observed light distribution with a constant stellar mass-to-light ratio, $M_\star/L$. The models also assume a central black hole of mass $M_{BH}$ and a cored isothermal dark-matter halo with core radius $r_{DM}$ and asymptotic circular velocity $v_{DM}$. The four parameters of the mass model are constrained by the photometrically derived luminosity density and by 1,978 LOSVD data points measured from the galaxy spectra between 0.4 arcsec and $\sim$45 arcsec.

Given some specific values for $M_\star/L$, $M_{BH}$, $r_{DM}$ and $v_{DM}$, the Poisson equation is solved for the gravitational potential generated by the respective mass model and the luminosity densities, and the LOSVDs of $\sim$29,000 representative stellar orbits are computed[9]. The orbits are convolved with the PSF of the observations and integrated over the respective areas on the sky. We use a maximum entropy technique[42] to determine the orbital occupation numbers that minimize the $\chi^2$ difference among the observational data and the orbit superposition model. Thousands of different mass distributions are compared to the data by systematically varying $M_\star/L$, $M_{BH}$, $r_{DM}$ and $v_{DM}$. We repeat the computation of the stellar orbits and the phase-space optimization independently for each model. The best-fit values and confidence intervals in $M_{BH}$ and $M_\star/L$ are determined by evaluating the relative likelihood[43] for all models with different assumed values of $M_{BH}$, $M_\star/L$ and dark halo parameters (Extended Data Fig. 3). The same modelling technique to determine the mass of the stars, the central black hole and the dark matter halo was first applied to the central galaxy of the Virgo Cluster, M87 (ref. 44).

Previous models of NGC 1600, based on stellar velocity data only along the major and minor axes of the galaxy and with a lower spatial resolution, did not include all three mass components[45–48]. We have tested that models without a dark-matter halo and/or without a central black hole cannot reproduce the full set of our new observations. Extended Data Fig. 4 shows our best-fit orbit model together with the velocity data for NGC 1600.

**Stellar mass-to-light ratio.** The stellar mass-to-light ratio derived from our dynamical modelling is $M_\star/L = (4.0 \pm 0.15) M_\odot/L_\odot$. The stellar mass deficit in the core of NGC 1600 is thus $\Delta M_{def} = 3.8 \times 10^{10} M_\odot$. In other core galaxies, mass deficits have been reported to range from one to ten times the mass of the central black hole[15,17,49]. For NGC 1600 we find a mass deficit, $\Delta M_{def}$, of $2.2 \times M_{BH}$. Results from numerical simulations of mergers of galaxies with central black holes[50] suggest mass deficits of $\sim N \times 0.5 M_{BH}$, where $N$ is the number of mergers.

We measure a stellar population age of $\tau \approx 10$ Gyr, a total metallicity of [Z/H] = 0.03 and an iron abundance of [Fe/H] = $-0.15$ for NGC 1600, from the absorption-line strengths of hydrogen, iron and other metallicity-dependent stellar lines in the 1-kpc galaxy core. It has been reported that the fraction of low-luminosity dwarf stars in elliptical galaxies less luminous than NGC 1600 is larger than that in the Milky Way[51–53]. Extrapolating correlations with the galaxy velocity dispersion obtained for these smaller elliptical galaxies would yield a higher stellar mass of $M_\star/L \approx 6.0$ for NGC 1600. We find the galaxy-wide contribution of dwarf stars to the stellar mass in NGC 1600 to be consistent with observations of the Milky Way. Dynamical constraints on the fractional mass of dwarf stars depend on the assumed dark-matter halo profile[54] and on the assumed shape of the stellar initial mass function[55]. Some nearby massive galaxies have, like NGC 1600, dwarf-star fractions consistent with that of the Milky Way[56].

The spectroscopic analysis of the age and chemical composition of the stars in NGC 1600 does not provide evidence for a notable change in the stellar population with radius. In the dynamical modelling we assume a constant $M_\star/L$ ratio throughout the galaxy. Individual massive galaxies have been reported to host extreme populations of dwarf stars at their centres[57] that would not be detectable in our optical spectra. The respective dwarf stars can increase the central $M_\star/L$ by up to a factor of three. If the assumed constant $M_\star/L$ does not account for all of the stellar mass at the centre of NGC 1600, then the dynamical models may compensate for the missing stellar mass by overestimating $M_{BH}$. Extended Data Fig. 5 shows the enclosed mass distribution of NGC 1600 over the region for which we have obtained stellar velocity data. For a constant $M_\star/L$, we find the enclosed stellar mass at the smallest observed radius ($r \approx 0.2$ kpc) to be 100 times smaller than $M_{BH}$. A central increase of $M_\star/L$ by a factor as extreme as ten would imply an unaccounted-for extra stellar mass of 10% of $M_{BH}$ (dotted lines in Extended Data Fig. 5). Even in this unrealistic case, we would overestimate $M_{BH}$ only by its one sigma measurement error. Because there is so little stellar light in the core of NGC 1600, uncertainties in the central stellar population have a negligible effect on our black-hole mass measurement.

**Distribution of stellar orbits.** The stars in a galaxy are collisionless and their velocity distribution can be anisotropic. We compute the intrinsic velocity dispersions of the stars along the radial and the two angular directions of a polar coordinate system—$\sigma_r$, $\sigma_\vartheta$, and $\sigma_\varphi$, respectively—from the orbital occupation numbers of our best-fit dynamical model in 20 spherical shells centred on NGC 1600's black hole. The classic measure for the anisotropy of the stellar velocities is $\beta = 1 - \sigma_t^2/\sigma_r^2$, where the tangential velocity dispersion $\sigma_t = [(\sigma_\vartheta^2 + \sigma_\varphi^2)/2]^{1/2}$ is the average of the motions in the two angular directions. Stellar orbits in core galaxies have been reported to be very uniform[18]. In NGC 1600 and similar galaxies with a flat central surface brightness, most of the stars inside the diffuse core region ($r < r_b$) are moving along tangential directions. With increasing distance from the centre, more and more stars are found on radially elongated orbits (Extended Data Fig. 6).

In the black-hole-binary model, the observed stellar motions are naturally explained as the leftover of the core scouring process (shaded regions in Extended Data Fig. 6). Central stars originally on radial trajectories are subject to interactions with the black-hole binary as they frequently pass the galaxy centre. Eventually, these stars get ejected to larger radii via gravitational slingshot. The stars that we observe today in the centres of core galaxies remained there because they moved (and still move) on tangential orbits that avoid the centre[58].

It has not yet been tested whether other black-hole activities—such as their feedback processes on ambient accreting gas—can produce the tight relations between black-hole mass, core radius, sphere of influence and mass deficit together with the observed orbital structure.

**Measures of galaxy core size.** While the core-Sérsic function describes galaxy light profiles from the core region out to large radii[30,59], the Nuker function[60] has been widely used to fit the central light profiles of galaxies observed with HST. Fifteen out of the 21 core galaxies discussed in the text also have core radii measured from Nuker fits[13]. For the Nuker $r_b$, we obtain $\log_{10}(r_{SOI}/\text{kpc}) = (-0.18 \pm 0.21) + (1.00 \pm 0.09)\log_{10}(r_b/\text{kpc})$ with an intrinsic scatter of $\epsilon = 0.16$, and $\log_{10}(M_{BH}/M_\odot) = (10.06 \pm 0.45) + (1.25 \pm 0.17)\log_{10}(r_b/\text{kpc})$ with an intrinsic scatter of $\epsilon = 0.31$. The Nuker $r_b$ values were measured along the major axis of the galaxies, while the core-Sérsic $r_b$ values discussed in the text come from the galaxies' circularized light profiles[15].

Galaxy core sizes have also been quantified by the cusp radius[61], $r_\gamma$—that is, the radius at which the negative logarithmic slope of the surface brightness profile equals 1/2. We obtained the cusp radii of the galaxies that are shown in Figs 3 and 4 from their core-Sérsic models. Using $r_\gamma$ as a measure of the core size, we find $\log_{10}(r_{SOI}/\text{kpc}) = (0.06 \pm 0.28) + (0.94 \pm 0.09)\log_{10}(r_\gamma/\text{kpc})$ (intrinsic scatter $\epsilon = 0.16$) and $\log_{10}(M_{BH}/M_\odot) = (10.37 \pm 0.60) + (1.20 \pm 0.19)\log_{10}(r_\gamma/\text{kpc})$, $\epsilon = 0.33$. Note that the core of NGC 1550 has a slope[15] of $\gamma = 0.52 \pm 0.05$. The scaling relations with $r_\gamma$ have been computed without including NGC 1550. The slope and the scatter in the above correlations are consistent with the results for the core-Sérsic $r_b$.

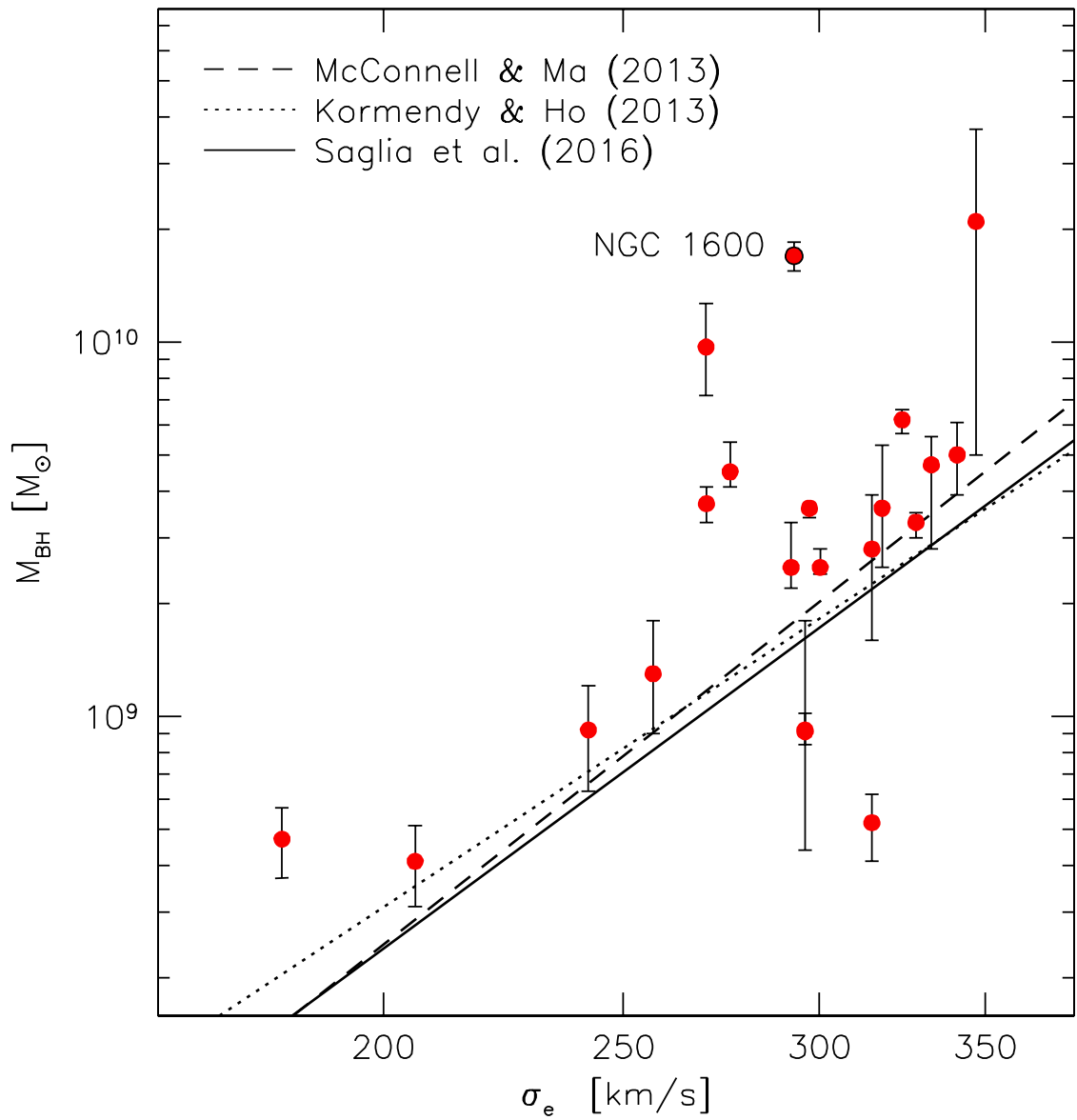

**Extended Data Figure 1 | The $M_{BH}/\sigma$ correlation.** The black-hole masses, $M_{BH}$, and host-galaxy velocity dispersions, $\sigma_e$, of the 21 core galaxies (red dots) shown in Figs 3 and 4. The dashed, dotted and solid lines show recent fits (from refs 10, 11 and 12, respectively) to the $M_{BH}$–$\sigma$ correlation for all early-type galaxies (including both cored and coreless galaxies) and classical bulges with dynamically measured $M_{BH}$. The black hole in NGC 1600 is ten times more massive than would be expected given the galaxy's velocity dispersion ($\sigma_e = 293$ km s$^{-1}$).

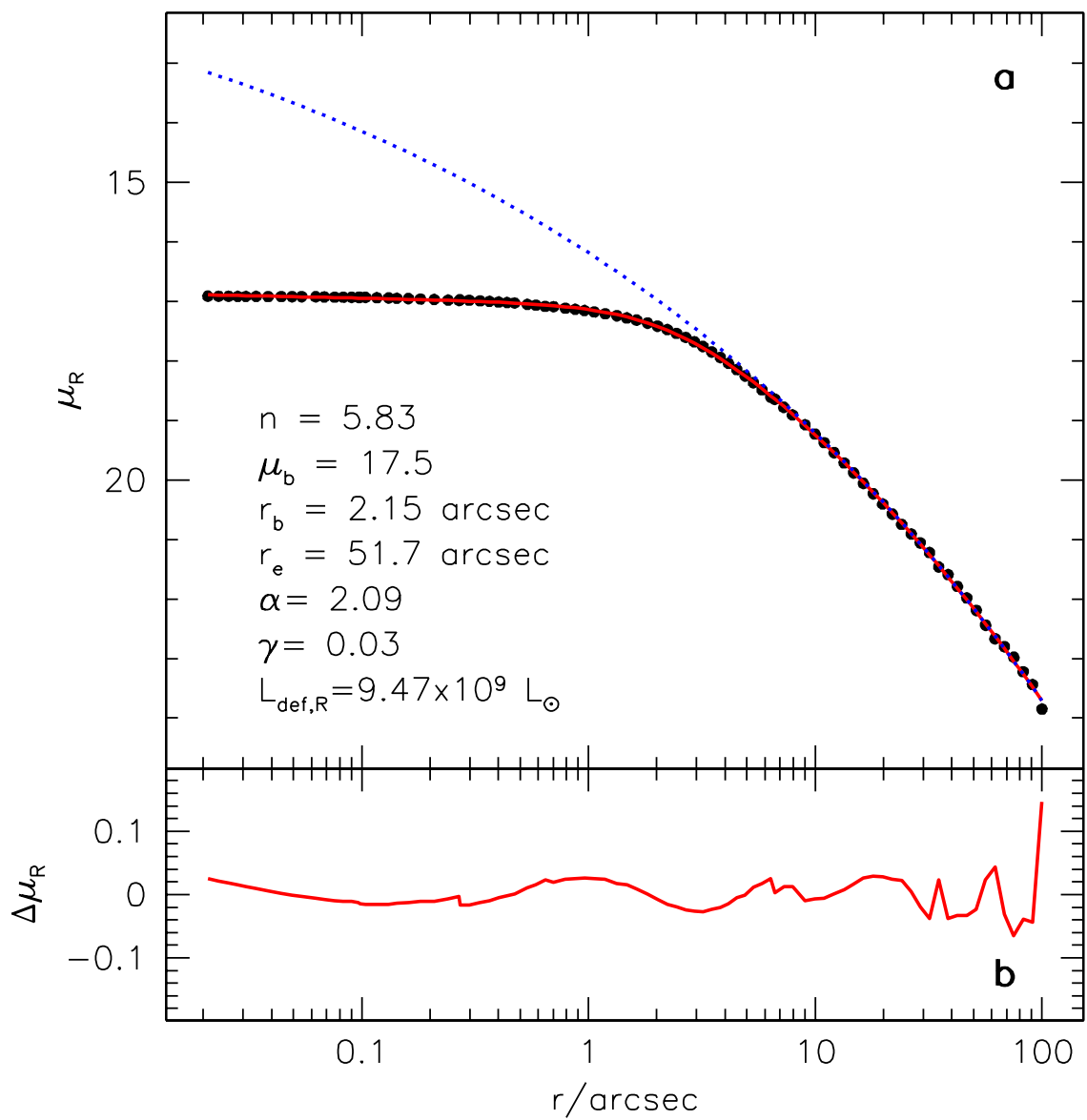

**Extended Data Figure 2 | Surface brightness profile of NGC 1600.**
**a**, The circularized surface brightness distribution of NGC 1600 (filled circles) and the best-fit core-Sérsic model (red line; the best-fit parameters of the core-Sérsic function are quoted). The blue dotted line indicates the inward extrapolation of the outer Sérsic component. From the integrated difference between the blue and the red curves, we derive a 'light deficit' of $L_{def} = 9.47 \times 10^9 L_\odot$. **b**, The difference between the data points in panel **a** and the core-Sérsic fit. Surface brightnesses are given in mag arcsec$^{-2}$ in the R-band.

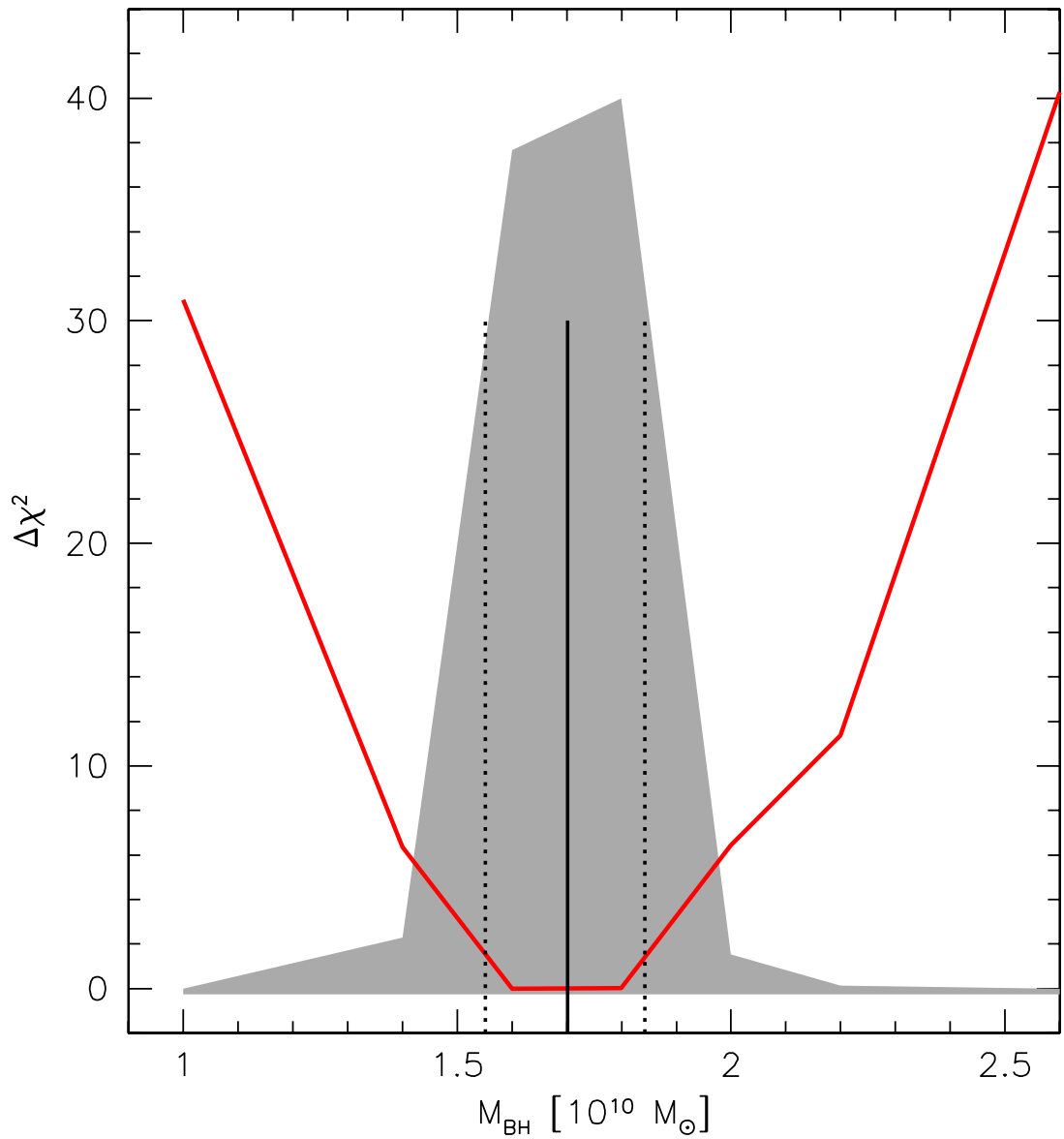

**Extended Data Figure 3 | Best-fit $M_{BH}$ values for NGC 1600 and confidence intervals.** The relative likelihood of different $M_{BH}$ values, marginalized over $M_\star/L$, $r_{DM}$ and $v_{DM}$ (shaded area; the likelihood is arbitrarily scaled). The best-fit values and confidence intervals are derived from the cumulative likelihood distribution[43] and indicated by the vertical lines. The red line shows $\Delta\chi^2(M_{BH}) = \chi^2(M_{BH}) - \chi_0^2$, where $\chi^2(M_{BH})$ is the minimum of all models with the same $M_{BH}$, but different $M_\star/L$, $r_{DM}$ and $v_{DM}$; $\chi_0^2$ is the minimum of $\chi^2(M_{BH})$ over $M_{BH}$.

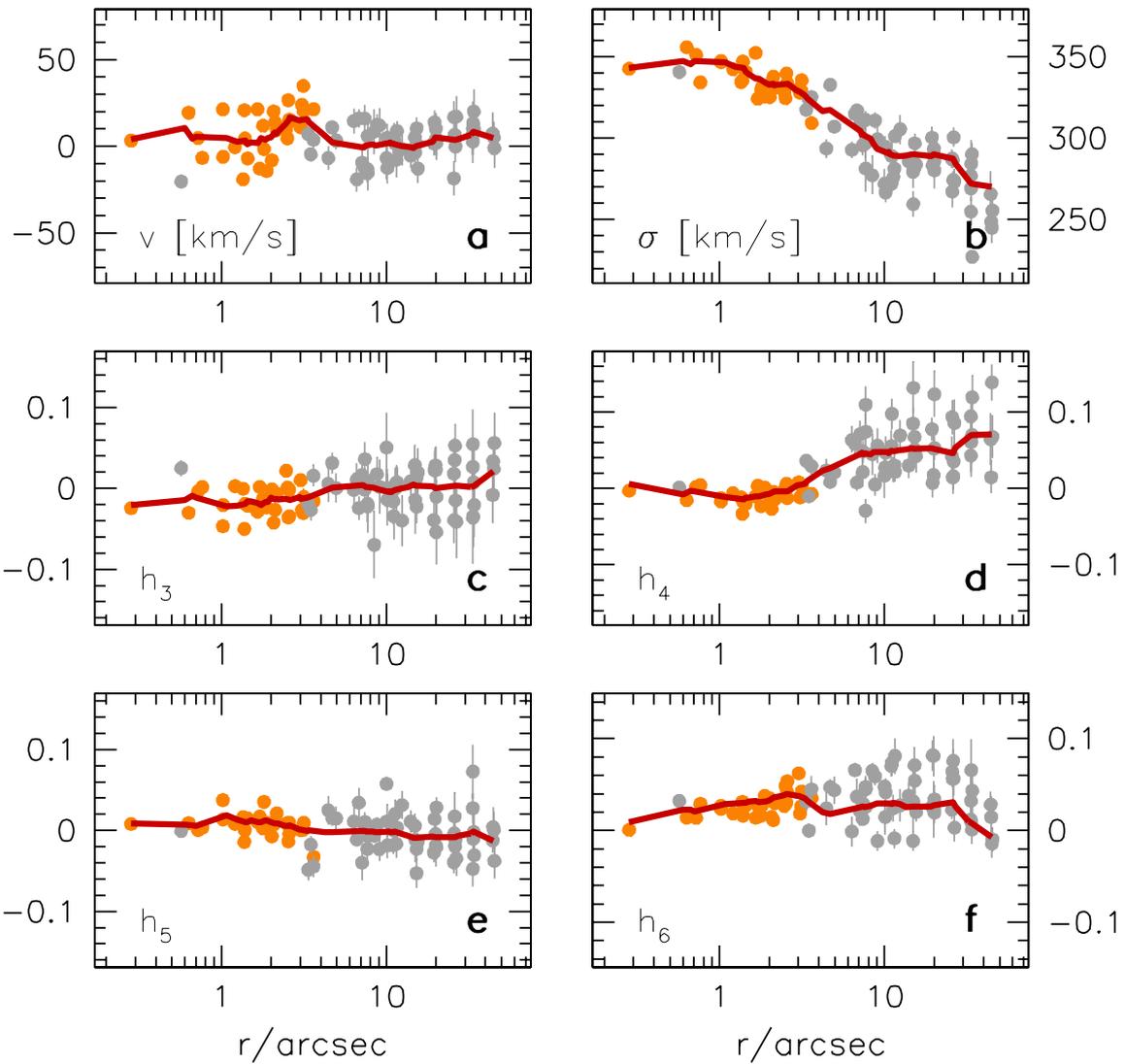

**Extended Data Figure 4 | Stellar velocity data and best-fit dynamical model.** These data are shown for NGC 1600 (filled grey and orange circles), together with the best-fit model (smoothed over 0.05 dex in log radius; solid red curves). Observed LOSVDs of galaxies are approximately Gaussian and are commonly parameterized by a Gauss–Hermite series expansion[62,63]. The mean stellar velocity $v$ (in **a**) and velocity dispersion $\sigma$ (in **b**) correspond to the centre and the width, respectively, of the best Gaussian approximation. Higher-order Hermite coefficients $h_n$ (in **c**–**f**) quantify deviations from a pure Gaussian LOSVD. Most data points at $r < 4$ arcsec came from our GMOS IFS observations (orange dots). Data at larger radii came from our Mitchell IFS observations (grey dots).

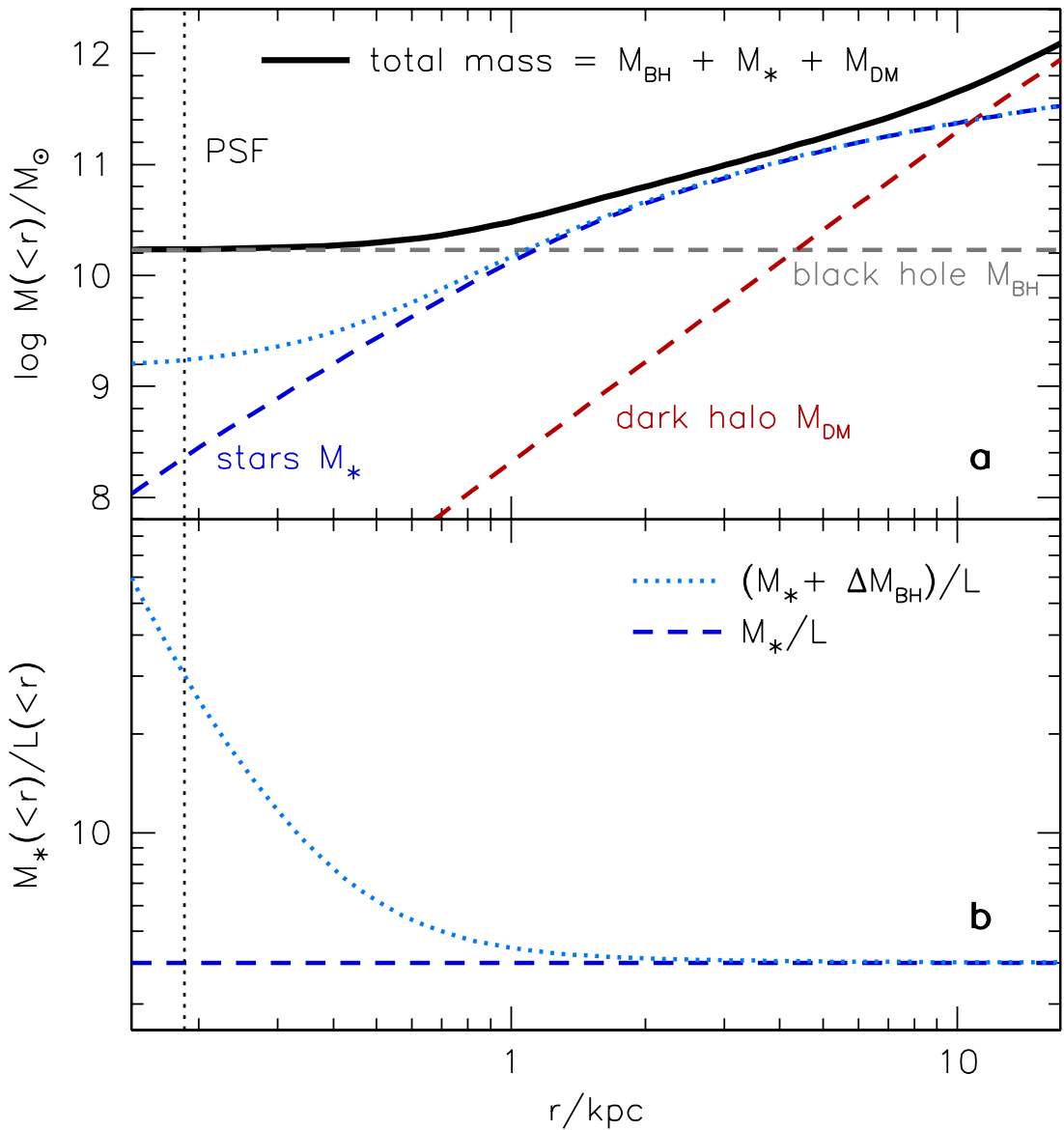

**Extended Data Figure 5 | The enclosed mass of NGC 1600. a**, The enclosed stellar mass ($M_\star$, blue), dark-halo mass ($M_{DM}$, red), black-hole mass ($M_{BH}$, grey) and combined total mass (black) obtained in our model from the smallest resolved radius (point-spread-function, PSF, size) out to 20 kpc (Mitchell IFU size). **b**, An illustration of the excessive $M_\star/L$ gradient (dotted pale blue curve) that would be required for a hypothetical population of unresolved central dwarf stars to explain 10% of NGC 1600's measured $M_{BH}$. The stellar mass-to-light ratio would have to increase by about a factor of ten (dotted pale blue curve) over our best-fit constant value (dashed blue curve). Observations of other galaxies suggest that extreme populations of dwarf stars can increase $M_\star/L$ by a factor of up to three.

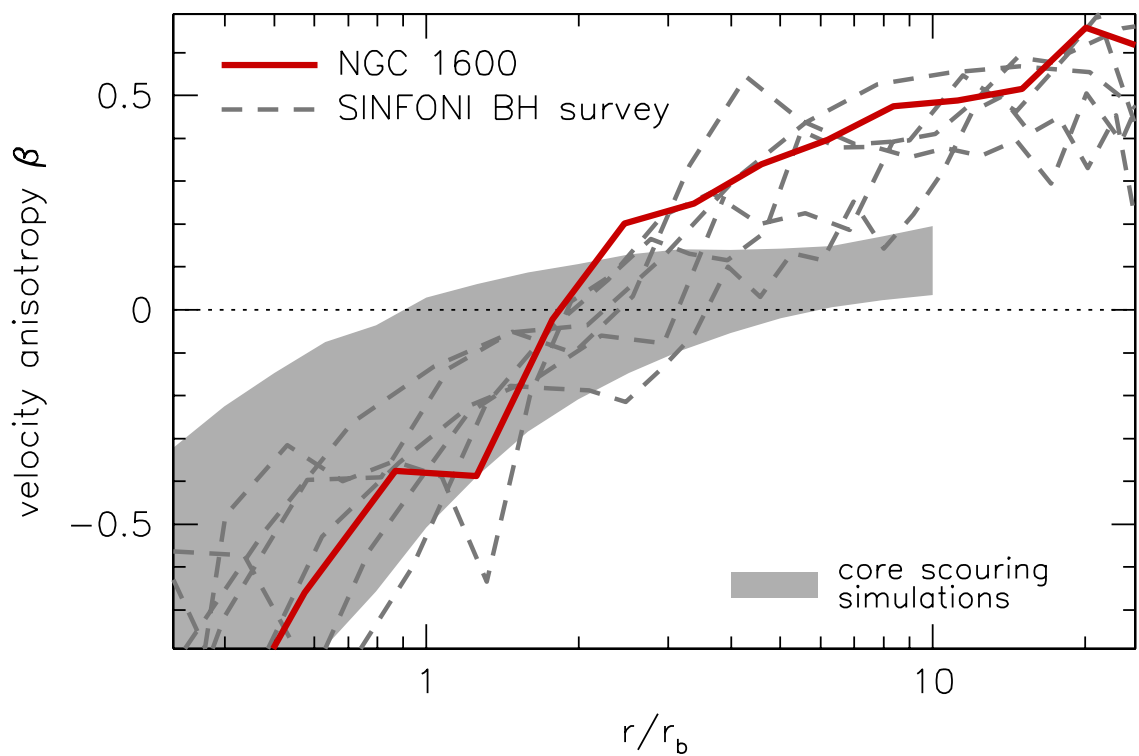

**Extended Data Figure 6 | The anisotropy of stellar orbits in core galaxies.** In NGC 1600 (red line) and similar galaxies with cores[12,18] (grey lines), the stellar velocity distribution is anisotropic. The anisotropy parameter, $\beta = 1 - \sigma_t^2/\sigma_r^2$, is positive when most of the stars move along radially stretched orbits, and negative when the stellar orbits are predominantly tangential. Inside the diffuse, low-surface-brightness core region ($r \leq r_b$), tangential motions dominate. The shaded area indicates the range of anisotropies found in numerical $N$-body simulations of the core scouring mechanism[58,64].